\documentclass[12pt]{article}

\usepackage[left=2.5cm,top=3.5cm,right=2.5cm,bottom=2cm,nohead]{geometry}

\usepackage{setspace}

\usepackage[english]{babel}
\usepackage[utf8]{inputenc}
\usepackage[T1]{fontenc}
\usepackage{type1ec}
\usepackage{graphicx} 
\usepackage{subfigurequickfix}
\usepackage{setspace}

\usepackage{listings}
\usepackage{parcolumns}
\usepackage{color}
\usepackage{amsfonts}
\usepackage{colortbl}
\usepackage{booktabs}
\usepackage{amsmath}
\usepackage{algorithmicx}
\usepackage{algorithm}
\usepackage{algpseudocode}
\usepackage{tabularx}
\usepackage{textcomp}
\usepackage{soul}
\usepackage{footmisc}
\usepackage{changepage}
\usepackage[dvipsnames]{xcolor}
 \usepackage[all]{nowidow}
  \usepackage{enumerate}

\newcommand{\aspas}[1]{{``#1''}}

\newcommand{\mcode}[1]{\texttt{#1}}

\definecolor{jsbackground}{RGB}{253,246,227}
\definecolor{jscomment}{RGB}{147,161,161}
\definecolor{jsstring}{RGB}{42,161,152}
\definecolor{jskeywords}{RGB}{133,153,0}
\definecolor{jsidentifier}{RGB}{88,110,117}
\definecolor{jsndkeywords}{RGB}{38,139,210}

\lstdefinelanguage{JavaScript}{
    keywords = {typeof, new, true, false, catch, return, null, catch, switch, if, in, while, do, else, case, break, +, <, >, *, /, =, ===, ==, +=, -=, *=, /=, <=, >=, !=, ++, --},
    ndkeywords = {function, var, this, prototype, Array, String, Number},
    sensitive = true,
    alsoletter = {<>+-=*/},
    alsodigit = {-},
    comment = [l]{//},
    morecomment = [s]{/*}{*/},
    morestring = [b]',
    morestring = [b]"
}

\lstdefinelanguage{HTML5}[]{HTML}{
    sensitive=false,
    alsoletter={-},
    morekeywords={section, ng-app, ng-controller, ng-submit, header, ng-model, placeholder, autofocus},
    tag=[s]
}

\lstdefinestyle{jsstyle}{
    backgroundcolor=\color{jsbackground},   
    commentstyle=\itshape\color{jscomment},
    stringstyle=\color{jsstring},
    keywordstyle=\color{jskeywords},
    ndkeywordstyle=\color{jsndkeywords}\bfseries,
    numberstyle=\tiny\color{jsstring},
    identifierstyle=\color{jsidentifier},
    basicstyle={\scriptsize\ttfamily},
    xleftmargin={0.75cm},
    numbers=left,
    stepnumber=1,
    firstnumber=1,
    numberfirstline=true,
    breakatwhitespace=false,
    breaklines=true,                 
    captionpos=b,                 
    keepspaces=true,                 
    numbers=left,                   
    numbersep=5pt,               
    showspaces=false,                
    showstringspaces=false,
    showtabs=false,                  
    tabsize=2,
    upquote=true
}

\begin{document}

\title{AngularJS Performance: A Survey Study}
\author{Miguel Ramos$^1$, Marco Tulio Valente$^1$, Ricardo Terra$^2$}
\date{\small $^1$Dept. of Computer Science, UFMG, Brazil \\
$^2$Dept. of Computer Science, UFLA, Brazil \\
\{miguel, mtov\}@dcc.ufmg.br, terra@dcc.ufla.br}
\maketitle

\begin{abstract}
\noindent AngularJS is a popular JavaScript MVC-based framework to construct single-page web applications.
In this paper, we report the results of a survey with 95 professional developers about performance issues of AngularJS applications.  We report common practices followed by developers to avoid performance problems (e.g., use of third-party or custom components), the general causes of performance problems in AngularJS applications (e.g., inadequate architecture decisions taken by AngularJS users), and the technical and specific causes of performance problems (e.g.,~unnecessary processing included in the digest cycle, which is the internal computation that automatically updates the view with changes detected in the model).
\end{abstract}


\section{Introduction}
\label{chap:introduction}

\begin{figure}[htb]
\centering
\def\svgwidth{0.8\textwidth}
\input{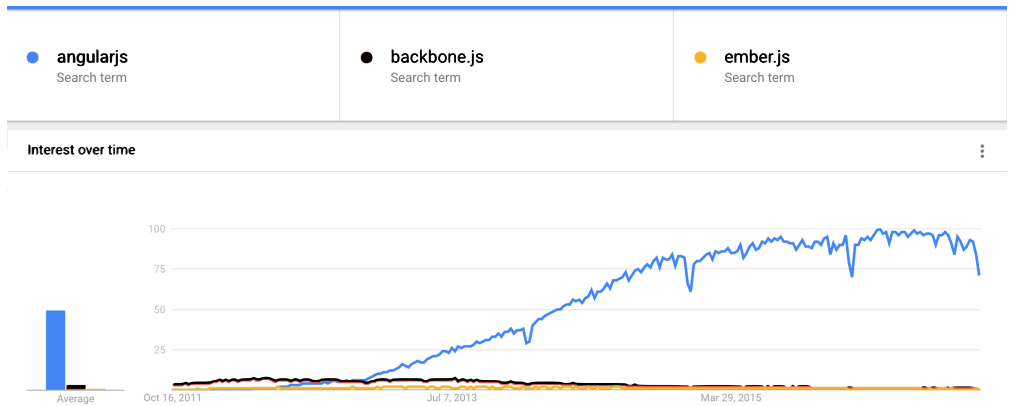}
\caption{Google searches for AngularJS, Backbone, and Ember}
\label{fig:googletrends}
\end{figure}

\begin{figure}[htb]
\centering
\def\svgwidth{0.8\textwidth}
\input{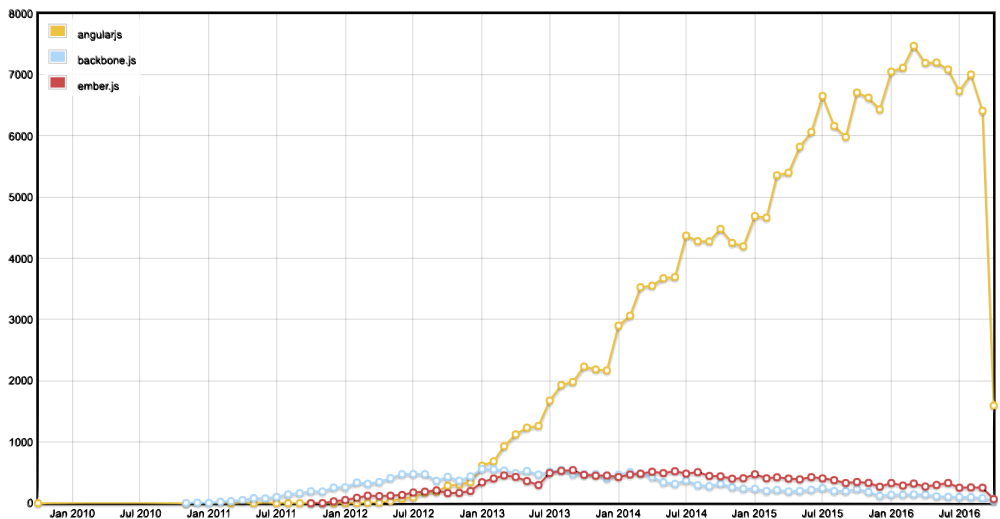}
\caption{Number of Stack Overflow questions per week for AngularJS, Backbone, and Ember}
\label{fig:sotrends}
\end{figure}

JavaScript is a fundamental piece of modern Web applications. The language is used nowadays to construct a variety of systems, including Web applications with sophisticated user interfaces. As a result, we are observing the birth of new technologies and tools (including JavaScript libraries and frameworks) to solve common problems faced in such applications. 
Specifically, a new family of JavaScript frameworks emerged, following the Model-View-Controller (MVC) architecture pattern (or variations of it). As examples, we have AngularJS, Backbone.js, and Ember.js. Among them, AngularJS~\cite{angularjsdoc,2016_plateau}, which is built and maintained by Google, is the most popular framework. This fact is evidenced by comparing the number of Google searches (the most queried framework since 2013, as illustrated in Figure~\ref{fig:googletrends}), the number of questions in Stack Overflow (the one with more Q\&A per month since 2013, as illustrated in Figure~\ref{fig:sotrends}),
and the numbers of contributors/stars at GitHub (1,530/52,718 for AngularJS against 620/16,938 for Ember.js, and 289/25,615 for Backbone.js), as collected on October 2016.
Finally, as examples of well-known websites using AngularJS, we have \url{http://weather.com}, \url{http://forbes.com}, and \url{http://intel.com} (according to \url{http://libscore.com}, a service that collects stats on JavaScript library usage).

Despite the increasing interest on AngularJS, there are few studies about  the performance of the applications constructed with the framework, including the recurrent performance problems faced by AngularJS users, and the possible causes and solutions to them. To contribute with a set of best practices to deal with performance problems in AngularJS, this paper reports the results of a survey about this specific aspect of the framework. We also reveal the design decisions and features of the framework that are usually seen as sources of performance problems. 

\section{Survey Design}
\label{sec:angulardesign}

First, we use a mapping study to get information about the main challenges and problems faced by developers, regarding the performance of  AngularJS.  A mapping study is more flexible than a systematic literature review. For this reason, it is recommended  for studying emergent fields or technologies~\cite{wohlin2012experimentation}, like AngularJS. We focus on blogs, forums, and sites of questions and answers (e.g.,~Stack Overflow), since there is limited formal literature on AngularJS. We start by using Google search queries such as \aspas{\emph{AngularJS performance}} or \aspas{\emph{performance problems in AngularJS}} to collect the first documents of interest. We review these documents to check
whether they indeed discuss performance aspects of AngularJS. 
For frequently mentioned topics, we perform new searches to find more information. As an example, we query for \aspas{\emph{ng-repeat performance}} because discussions about this directive appear very often in the initial set of reviewed documents.
In total, we collect 25 documents (3 AngularJS documents, 14 blog posts, and 8 Stack Overflow questions). We review each document to identify common trends and discussions, which were used in the survey design.

The survey has 31 questions divided in four sections: (1) background, (2) practices and perceptions w.r.t.~performance in AngularJS, (3) general causes of AngularJS performance problems, and (4) technical causes of these problems. 
The questions regarding the technical causes of performance problems are not mandatory because they may require knowledge on advanced features of the framework. Therefore, participants are not forced to provide an answer when they did not master an specific feature responsible for  performance issues. 

We promote the survey in popular AngularJS communities and forums, including the AngularJS Google group\footnote{\url{https://groups.google.com/forum/\#!forum/angular}}, the AngularJS community in Google+\footnote{\url{https://plus.google.com/communities/115368820700870330756}}, and a Reddit community\footnote{\url{https://www.reddit.com/r/angularjs/}}. The survey remained open during three weeks (since early September 2015) and we obtained 95 responses. The documents used in the mapping study and the survey questionnaire are publicly available in a companion repository.\footnote{\url{https://github.com/aserg-ufmg/angularjs-performance-survey}}

\section{Results}
\label{sec:angularperf_results}

\subsection{Background}
\label{subsec:backgroundperf}

\begin{figure}[htpb]
\centering
\centering
\subfigure[JavaScript experience \label{fig:jsperfexperience}]{
\includegraphics[width=0.31\linewidth]{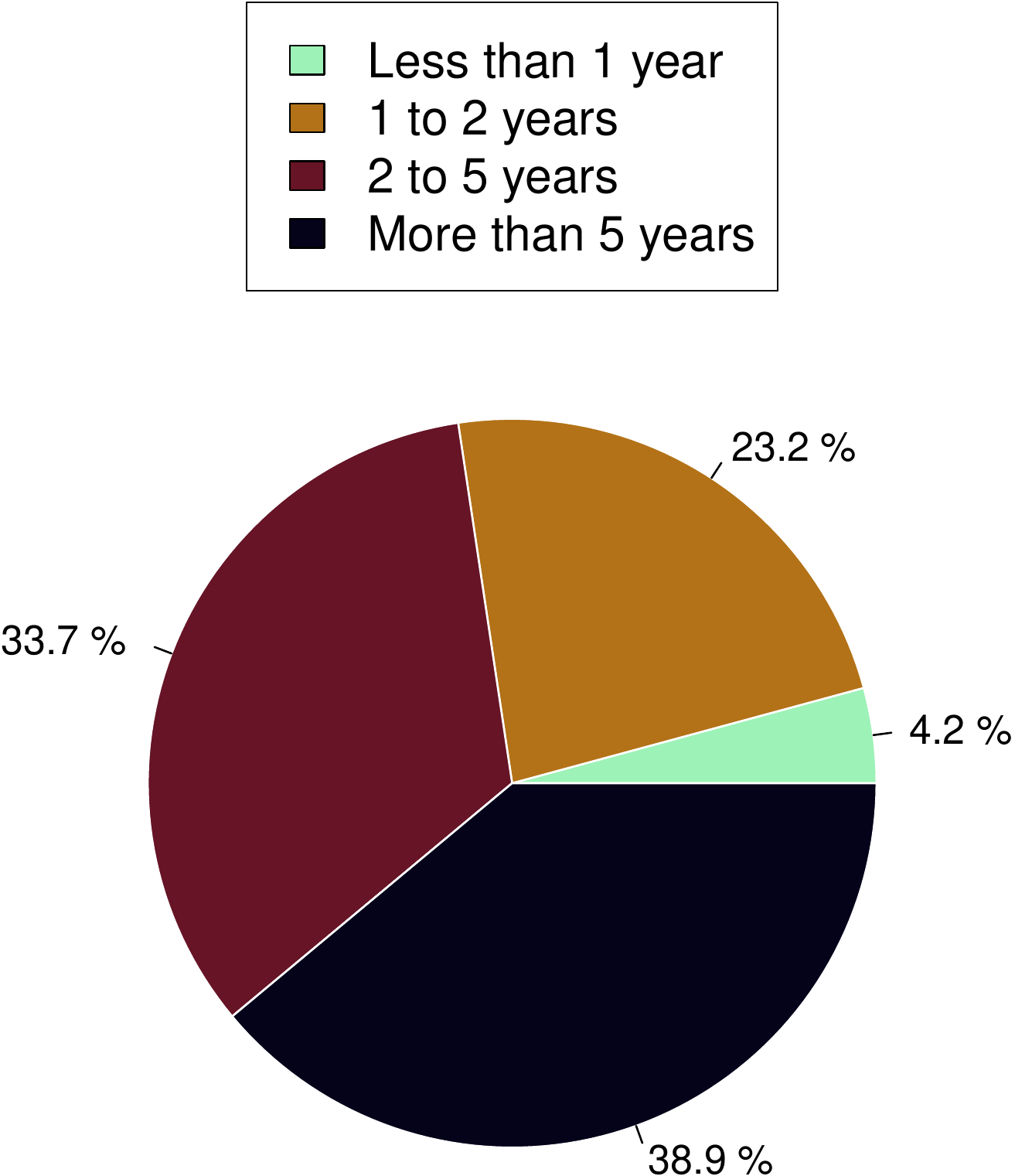}}
\subfigure[AngularJS experience \label{fig:angularperfexperience}]{
\includegraphics[width=0.31\linewidth]{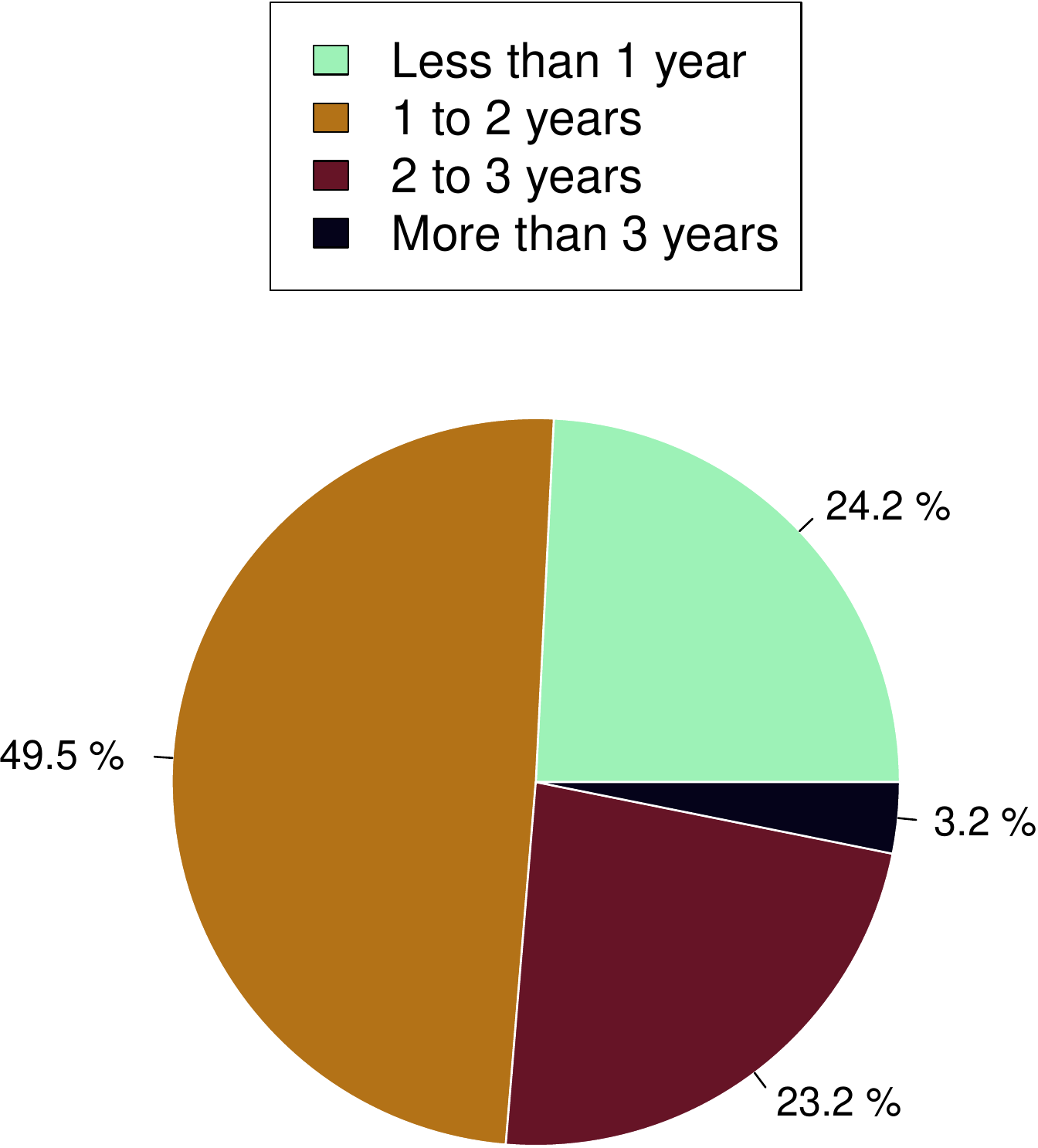}}
\subfigure[Largest application \label{fig:largestperfapplication}]{
\includegraphics[width=0.31\linewidth]{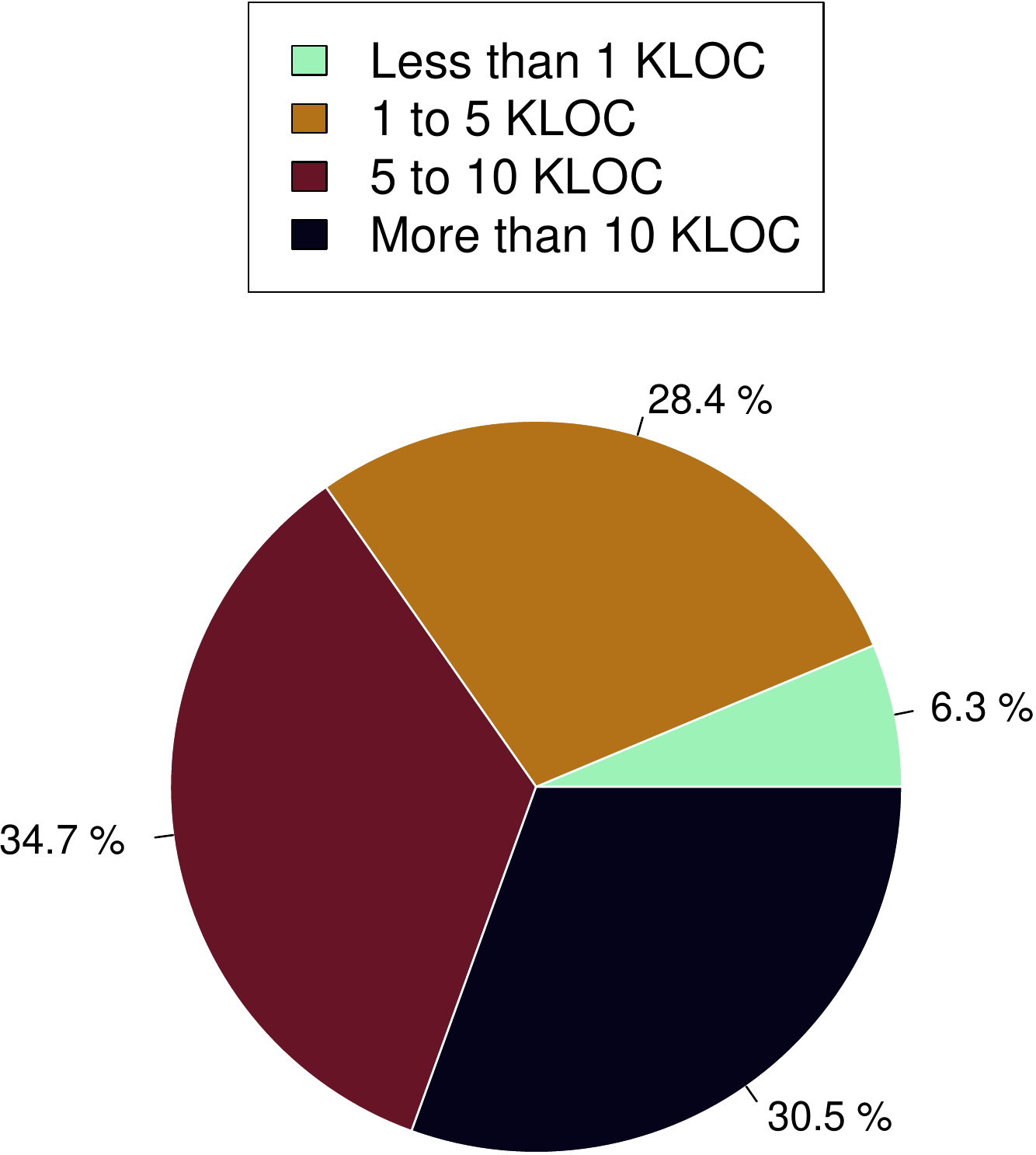}}
\caption{Participants background}
\label{fig:backgroundperf}
\end{figure}

As presented in Figure~\ref{fig:backgroundperf}, 72.6\% of the respondents have at least two years of experience in JavaScript and 75.8\% have at least one year of experience in AngularJS. Only 3.2\%  have more than 3 years of experience in AngularJS since the framework started to become popular around 2013 (although its first release is from 2009). Only 6.3\% of the participants report that their largest AngularJS application  has less than 1 KLOC. Therefore, we can conclude that most survey participants are not novice AngularJS developers.

\subsection{How Developers Improve AngularJS Performance?}
\label{subsec:actionstoimproveperf}

\noindent{\em Practices:} Some questions reveal the practices followed by developers to improve the performance of their AngularJS applications as follows:
\begin{itemize}
\item 45.5\% of the respondents have at least once inspected the AngularJS source code to figure out how to improve the performance of their applications;
\item 8.4\% of the respondents changed the AngularJS source code to improve performance;
\item 29.5\% of the respondents use third-party components to improve performance; and
\item 27.4\%  of the respondents created a custom component to this purpose. 
\end{itemize}

\noindent{\em Usage of custom or third-party components:} 
According to the participants, the two most common  reasons for using custom or third-party components are:
\begin{itemize}
\item to support bind-once when this feature was not available  (before version~1.3); and
\item to replace or improve the standard \mcode{ng-repeat} directive.
\end{itemize}
In the first case, the only component mentioned is {\tt bind-once} while in the second case the following components are mentioned: {\tt angular-virtual-scroll}, {\tt infinite-scroll} (which is no longer maintained), {\tt ng-table}, and {\tt{collection-repeat}} from the Ionic framework. Mostly, these components deal with large amounts of data in tables or lists, which confirms a topic that emerged in the mapping study (see  two Stack Overflow posts~\cite{kashifmustafa,SO1}). \\

\begin{figure}[!t]
\centering
\vspace{0.5cm}
\includegraphics[width=0.97\linewidth]{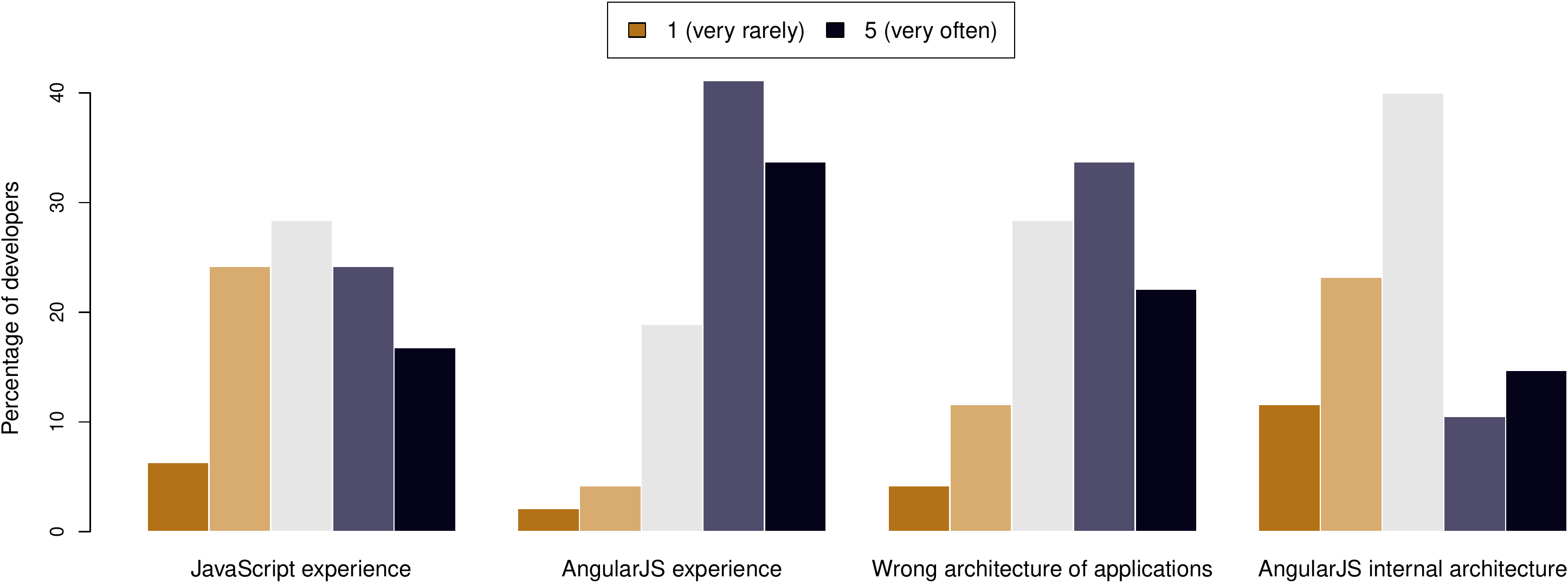}
\vspace{-0.1cm}
\caption{General causes of performance problems in AngularJS}
\label{fig:generalperfcauses}
\end{figure}

\noindent{\em Usage on Mobile phones:} 
In the mapping study, we found documents reporting that the performance problems in AngularJS are more critical in mobile phones~\cite{egorkoshelko,peterkoch}. When asked about this fact, 34 developers (35.8\%) answered that at least once they refactored their code to  improve performance specifically in mobile devices.

\subsection{General Causes of Performance Problems in AngularJS}
\label{sub:generalcausesperf}


Figure~\ref{fig:generalperfcauses} presents the answers of the developers from
survey questions about the following four possible general reasons for performance problems in a scale from one (very rarely) to five (very often):\\

\noindent {\em Lack of experience in JavaScript}: The most common answer is the neutral one (score~3) with 28.4\%. The number of answers with scores 2 and 4 is the same (23 answers) and the difference between the extreme options (scores 1 and 5) does not allow a clear conclusion.\\ 

\noindent {\em Lack of experience in AngularJS}: 74.8\% of the respondents  indicate the lack of experience in AngularJS as a frequent cause of performance problems (scores 4 and 5). \\

\noindent {\em Inadequate architectures}: 55.8\% of the developers indicate the adoption of inadequate architecture decisions as a general cause of performance problems. For instance, not using a caching when a large list of items needs to be computed each time it is presented to users.\\

\noindent {\em AngularJS architecture}: We check whether the internal architecture of AngularJS is the general cause of poor performance. In this case, most answers are neutral (40\%) and there is not a clear trend in either side of the scale.





\subsection{Technical Causes of Performance Problems}
\label{sub:technicalcausesperf}

The mapping study revealed 13 technical causes for performance problems in AngularJS:

\begin{enumerate}[({Cause}~\#1)]
\item {\em Unnecessary use of two-way data binding}, e.g., unnecessary synchronization between data in the view and in the model.
\item {\em Use of wrong watching strategies}. 
For example, during the digest cycle, values of the expressions being watched might be compared by collection or by value (instead of by reference). On one hand, this
%
allows the comparison of nested values in objects. On the other hand, it requires a full traversal of the nested data structure on each digest; moreover, a full copy of the data structure needs to be held in main memory~\cite{DO1}.
%
%
%
\item {\em Watching complex functions in the digest cycle}.  If complex functions are registered for watching during the digest cycle, performance can drop considerably.
\item {\em Use of filters in templates} that are evaluated on each digest cycle. 
\item {\em Use of mouse events directives}, which can trigger the digest cycle very often. 
\item {\em Use of a wrong conditional-display strategy}, e.g., \mcode{ngIf} and \mcode{ngSwitch} directives actually create or destroy DOM elements depending on a condition, which might be expensive operations.
\item {\em Inappropriate use of \mcode{\$scope.\$apply()}}, which is the method used by AngularJS to trigger the full digest cycle.
\item  {\em Runtime configurations and flags}, which should be disabled (but are not) when the application is in production. 
\item {\em Not cleaning up resources}, including handlers, watches, and asynchronous operations, even when they are no longer needed.
\item {\em Use of \mcode{ngRepeat} with long lists} rather than using pagination or infinite scroll. 


\item {\em Use of \mcode{ngRepeat} without \aspas{track by}}. This expression
avoids expensive operations---such as creation and deletion of DOM elements---by specifying
a unique key to each item to make the identification.
%
%

\item {\em Use of nested directives in \mcode{ngRepeat}}. It occurs when items of a collection used in a \mcode{ngRepeat} are represented by templates that create additional  bindings.
For instance, if four directives are nested in  a \mcode{ngRepeat} that is iterated 100 times, it would generate 400 new expressions that need to be watched during the digest~cycle.  
%

\item {\em Unconscious triggering of the full digest cycle}. Some
methods that trigger the full digest cycle are
 \mcode{\$http}, \mcode{\$resource}, and \mcode{\$timeout}. 
The lack of knowledge of this
fact can lead developers to architect applications where the digest
cycle is triggered multiple times.

\end{enumerate}

Most causes are related to the digest cycle.
AngularJS provides the ability to constantly maintain in sync the state of the application 
with the view presented to the final user. 
This synchronization is done by constantly observing the variables in the model.
Therefore, the more data to sync or 
the more complex are the watched functions,
the slower is this cycle.

\begin{figure}[!t]
\centering
\vspace{0.5cm}
\includegraphics[width=0.99\linewidth]{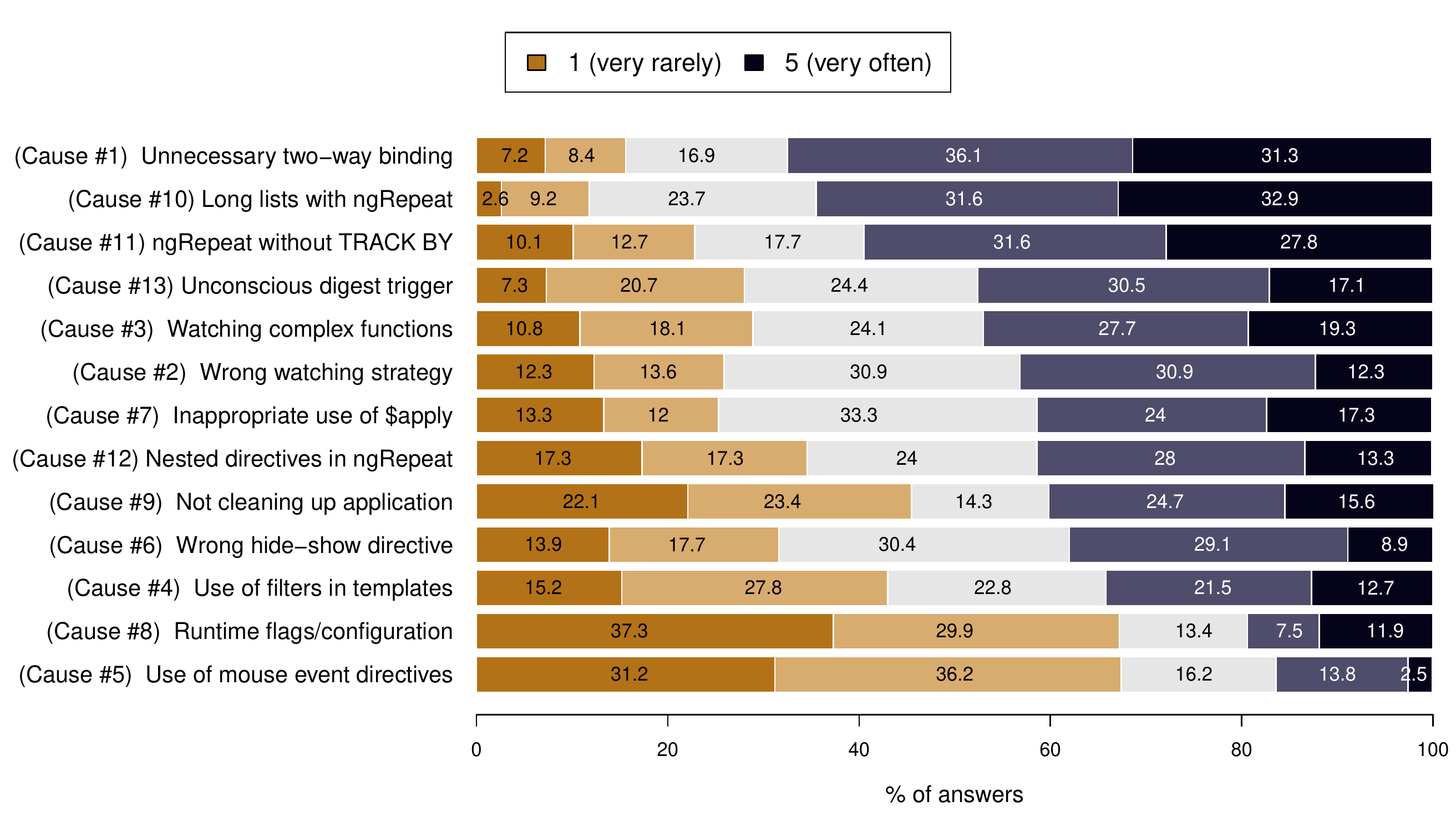}
\vspace{-0.1cm}
\caption{Technical causes of performance problems, ordered by \% of answers with scores 4-5}
\label{fig:technicalperfcauses}
\end{figure}

We ask the developers to rate how often these items are responsible for negative effects in performance. Figure~\ref{fig:technicalperfcauses} shows the results, in a scale from 1 (very rarely) to 5 (very often). 
The number of responses ranges from 67 (70.5\%, Cause~\#8) to 83  (87.4\%, Causes~\#1~and~\#3).

We also perform statistical tests over the data in Figure~\ref{fig:technicalperfcauses}. The answers from every cause deviate from normality using Shapiro-Wilk test (p-value ranged from $2.888e^{-5}$ to $7.162e^{-8}$).
Therefore, we use a Wilcoxon Signed-Rank Test with a 95\%
confidence interval
and the following null hypothesis: the median of the answers is less or
equal to 3. We rejected this null hypothesis for the first five causes in Figure~\ref{fig:technicalperfcauses}, 
i.e.,~Causes \#1, \#10, \#11, \#13, and \#3. Therefore, we have statistical
evidences that these items are often or very often viewed as causes
of performance problems.
Moreover, we executed the test again, with a second null hypothesis:
the median of the answers is greater or equal to 3.
In this case, it was rejected for the last two causes in Figure~\ref{fig:technicalperfcauses}, 
i.e.,~Causes \#8 and \#5. For these
causes we can state, with statistical support, that they are rarely or
very rarely
associated to performance concerns.

We first discuss the five causes with statistical evidences  they are often or very often responsible for
performance problems:
 
\begin{itemize}

\item Cause~\#1 (unnecessary use of two-way data binding) suggests that the easiness of keeping synchronized the data between the view and the model---in terms of no need to write any complex code---makes developers likely to overuse such feature.

\item Cause~\#10 (use of long lists with the \mcode{ngRepeat} directive) shows that developers agree that performance deteriorates when rendering long collections with the \mcode{ngRepeat} directive. This is a consistent topic addressed in the documents found in our mapping study and it is one of the problems that the AngularJS community attempts to solve through third-party components. Moreover, \mcode{ngRepeat}  can be easily used to create complex DOM elements for each item in a collection, creating at the same time several bindings (watches) that slow down the digest cycle. A common solution to this problem includes the reduction of the displayed list by means of pagination or infinite scroll~\mbox{\cite{chandermani,sebastianfrostl}}.

\item Cause~\#11 (use of the \mcode{ngRepeat} without the \emph{track by} expression) reveals how much attention must be put in the architecture of AngularJS applications when working with large amounts of data. Specifically, the \emph{track by} expression avoids expensive operations, such as creation and deletion of DOM elements. This happens by defining a global identifier that allows the direct association of items in the server with the elements rendered in the client. 

\item Cause~\#13 (unconscious triggering of the full digest cycle) relates to specific methods that by their own nature trigger the digest cycle, such as  \mcode{\$http}, \mcode{\$resource}, \mcode{\$timeout},
\mcode{\$interval}, \mcode{\$scope.\$apply}, etc. The answers suggest that it is not trivial to track all the possible execution scenarios where these methods are used.

\item Cause~\#3 (watching complex functions) happens when developers register, through the \mcode{\$watch} method, a function whose returned value is watched.  In each digest cycle, the return values of the last and current executions are compared. If the function is very complex, its execution will take longer, as well as the digest cycle.

\end{itemize}


By contrast, Cause~\#8 (use of wrong runtime configurations) and Cause~\#5 (use of mouse events directives) are not seen as related with performance degradation.
Regarding Cause~\#8, disabling debug information is particularly an action that is not seen by the respondents as a  performance booster. Moreover, the addition of information (scopes and CSS classes) to DOM elements is only perceived while the application is loaded for the first time. It has little or no effect in the digest cycle and for this reason it might have been considered as having a minor improvement in performance. In the mapping study, we also found developers reporting that no improvements are perceived after disabling debug data~\cite{SO2}.
In the case of mouse-events directives, the reason may be that the problem appears only when they are spread all over the application creating multiple potential triggers of the entire digest cycle.\\

\noindent{\em Overlapping results}: 
We run Spearman's Rho test to measure the strength of association between the participants' answers for pairs of causes. 
For each pair of causes~\mbox{($C_i$, $C_j$)}, we compute $\mathit{rho}(A_i, A_j)$, where  $A_i$ and $A_j$ are respectively vectors with all answers (in a scale from 1 to 5) that were given by the participants to causes $C_i$ and $C_j$. We only found a {\em moderate} correlation ($\mathit{rho}=0.557$) between 
Cause~\#2 (the use of wrong watching strategy)
and
Cause~\#3 (watching complex functions).
This finding is somehow expected since both causes enclose the \mbox{{\em watching}~concept}.


\subsection{Threats to Validity}
\label{sec:threatsinperf}

The first threat to validity is related to our mapping study since there is vast material on the Web about AngularJS and therefore we may have missed important documents. In the construction of a survey, there is always the risk of having ambiguous or unclear questions. We made our best to avoid these questions by using multiple iterations of small reviews when formulating the proposed questions. 
Finally, the survey participants might not be representative
of the general population of AngularJS developers (external validity). To mitigate
this threat, we advertised the survey in different online communities. 

\section{Concluding Remarks}
\label{sec:finalremarksperf}

Our main findings are as follows  
(i)~many developers (45.3\%) reported to have inspected the code of AngularJS to fix or understand performance issues;
(ii)~the most frequent causes (from a general point of view) of poor performance are the lack of knowledge in AngularJS and 
developers taking inadequate architectures decisions;
(iii)~the most frequent causes of performance degradation in AngularJS applications are related with the unnecessary use of two-way data binding and the incorrect use of the \mcode{ngRepeat} directive; and
(iv)~using mouse-events directives and not using the adequate runtime configurations are not considered as major threats to performance. As future work, we plan to conduct interviews with AngularJS developers to explore and reinforce these findings.

A new version of AngularJS (called Angular 2.0) was released recently. However, at least for a time, the two versions will be maintained in parallel, since Angular 2.0 is  not backward compatible with AngularJS 1.x. The most important architectural modification in Angular~2.0 is the new change detection mechanism, which uses immutable and observable data structures to detect changes made in the model. Specifically, immutable objects reduce the number of checks when comparing for equality complex and nested data structures. Observables also contribute to performance gains by providing specific events other objects can subscribe to in order to detect changes.
In future work, experiments can be conducted with real and benchmark applications to
measure the gains of performance achieved with immutables and observables.
These experiments can also be extended to include other benchmarks, designed to validate the developers' perceptions reported in this paper.

\section*{Acknowledgments}
Our research is supported by FAPEMIG and CNPq.

\bibliographystyle{plain}
\bibliography{bibfile}

\end{document}